# High-Entropy Rare Earth Tetraborides

Mingde Qin, Qizhang Yan, Haoren Wang, Kenneth S. Vecchio, Jian Luo[*]

*Department of NanoEngineering; Program of Materials Science and Engineering, University of California San Diego, La Jolla, CA, 92093, USA*

## Abstract

Six high-entropy rare earth tetraborides of the tetragonal $UB_4$-prototyped structure have been successfully synthesized for the first time. The specimens are prepared from elemental precursors via high-energy ball mill and in-situ reactive spark plasma sintering. The sintered specimens are >98% in relative densities without detectable oxide impurities (albeit the presence of minor hexaborides in some compositions). No detectable secondary phase is observed in the composition $(Y_{0.2}Nd_{0.2}Sm_{0.2}Gd_{0.2}Tb_{0.2})B_4$, which is proven homogeneous at both microscale and nanoscale. The Vickers microhardness are determined to be ~13-15 GPa at a standard indentation load of 9.8 N. A scientifically interesting observation is represented by the anisotropic lattice distortion from the rule-of-mixture averages. This work expands the family of high-entropy ceramics via fabricating a new class of high-entropy borides with a unique tetragonal quasi-layered crystal structure.

---

[*] Corresponding author. E-mail address: jluo@alum.mit.edu (J. Luo).

**Highlights:**

- High-entropy rare earth tetraborides are synthesized for the first time.
- Sintered specimens are almost fully dense bulk pellets.
- Homogenous elemental distributions at both microscale and nanoscale.
- Anisotropic lattice distortion from the rule-of-mixture averages are observed.
- This study expands the family of high-entropy ceramics.

## 1. Introduction

Tremendous research effort has been made in the field of high-entropy alloys (HEAs) since the seminal reports from Yeh et al. [1] and Cantor et al. [2] in 2004. HEAs have been demonstrated to possess excellent, and even unexpected, mechanical and physical properties [3]. In the last five years, the research on the ceramic counterparts to HEAs, high-entropy ceramics (HECs), has seen an exponential growth. Specifically, various high-entropy oxides (including rocksalt [4], pervoskite [5], fluorite [6], pyrochlore [7], and spinel [8] oxides), borides [9-11], silicides [12], carbides [13-16], nitrides [17, 18], and fluorides [19] have been successfully fabricated. Single-phase high-entropy aluminides (intermetallic compounds) [20] have also been made, thereby bridging HEAs and HECs. Like their metallic counterparts [21, 22], it was proposed that HECs can be generalized to compositionally-complex ceramics (CCCs), where medium-entropy and non-equimolar compositions can sometimes outperform their high-entropy counterparts [7, 23-25].

For high-entropy borides (HEBs), Gild et al. first reported the successful synthesis of six single-phase high-entropy diborides of the hexagonal $AlB_2$ structure in 2016 [11]. Further studies have fabricated high-entropy hexaborides of the cubic $CaB_6$ (i.e. $LaB_6$) structure [26, 27], high-entropy monoborides of the orthorhombic CrB structure [10], and high-entropy $M_3B_4$ borides [9]. Moreover, improved fabrication routes have been developed to make HEBs via boro/carbothermal reduction of metal oxides [28-30] and reactive sintering of elemental precursors [31-33]. Notably, Qin et al. recently synthesized single-phase high-entropy diborides and monoborides via in-situ metal-boron reactive spark plasma sintering (SPS) [10, 33]. The adoption of metal and boron precursors enables the flexibility in controlling the metal-to-boron ratio, thereby providing a generic route to explore the synthesis of HEBs of different crystal structures and stoichiometric ratios.

Rare earth metal borides can possess different crystal structures with various boron contents, including diborides, tetraborides, hexaborides, dodecaborides, and hectoborides of the $AlB_2$, $UB_4$, $CaB_6$, $UB_{12}$, and $YB_{66}$ prototype structures, respectively [34]. Among them, $UB_4$-prototyped tetragonal tetraborides (Space group: P4/mbm, No. 127) represent a prevalent crystal structure, including the reported formation of $YB_4$, $LaB_4$, $CeB_4$, $PrB_4$, $NdB_4$, $SmB_4$, $GdB_4$, $TbB_4$, $DyB_4$, $HoB_4$, $ErB_4$, $TmB_4$, $YbB_4$, and $LuB_4$ [34]. The lattice parameters and cation radii of these binary rare earth tetraborides (with the exception of YB$_4$) decrease monotonously with the increasing atomic number; from LaB$_4$ to LuB$_4$, the lattice parameter *a* decreases from 7.324 Å to 7.036 Å, and the lattice parameter *c* decreases from 4.181 Å to 3.974 Å, as the cation radius *r* decreases from 1.13 Å to 0.90 Å [34-36]. Note that this UB$_4$-prototyped rare earth tetraboride structure is different from orthorhombic CrB$_4$ (Immm and Pnnm), orthorhombic MgB$_4$ (Pnma), monoclinic MnB$_4$ (C2/m and P2$_1$/c), and the hexagonal WB$_4$ (P6$_3$/mmc) [37]. UB$_4$-prototyped rare earth tetraborides often exhibit interesting magnetic properties. While most of these rare earth tetraborides are paramagnetic [36, 38], YB$_4$, LaB$_4$, and LuB$_4$ are reported to be diamagnetic [39]. Furthermore, PrB$_4$ and TmB$_4$ exhibit anisotropy of their magnetic properties [40].

It is scientifically interesting to explore whether UB$_4$-prototyped high-entropy tetraborides (HETBs) can be made to further expand the family of HECs to include a new and unique (highly anisotropic, tetragonal, and quasi-layered, as discussed below) crystal structure. A schematic illustration of this UB$_4$-prototyped high-entropy tetraboride (HETB) structure can be found in Fig. 1. It can be interpreted as a hybrid of alternating MB$_2$ and MB$_6$ units at a 1:1 ratio along the (001) plane, where the octahedra composed of six boron atoms (in MB$_6$ units) are linked by boron-boron bonds to linear chains along the *c* axis, similar to the cubic axes of CaB$_6$-prototyped hexaborides. However, these MB$_2$ and MB$_6$ units in the UB$_4$-prototyped tetraboride are not exactly the unit cells of AlB$_2$ and CaB$_6$ types, but rather distorted.

Notably, this tetragonal HETB crystal structure (Fig. 1) is "quasi-layered" with alternating covalently-bonded boron layers and "2D high-entropy" metal layers (Fig. 1(b)), thereby being highly anisotropic. However, unlike high-entropy diborides of the hexagonal AlB$_2$ structure [11], the boron layers are still linked by linear boron-boron chains along the *c* axis at a low density (Fig. 1(b)), and high-entropy metals are "isolated" from one another ("caged" by covalently-bonded boron network). In this study, we adopt the in-situ metal-boron reactive SPS method [10, 33] to fabricate this new class of tetragonal HETBs that contain five rare earth metals selected from Y, La, Pr, Nd, Sm, Gd, Tb, Dy, Er, and Yb.

## 2. Experimental Procedure

Rare earth based HETB specimens were synthesized via in-situ metal-boron reactive SPS [10, 33]. Commercial elemental powders of Y, La, Pr, Nd, Sm, Gd, Tb, Dy, Er, and Yb (99.9% purity, ~40 mesh, purchased from Alfa Aesar, MA, USA) and boron (99% purity, 1-2 μm, purchased from US Research Nanomaterials, TX, USA) were utilized as precursors for making specimens of six desired compositions (HETB1 to HETB6) listed in Table 1. To synthesize each specimen, stoichiometric amounts of metals and boron powders were weighed out in batches of 5 g. The powders were initially mixed by a vortex mixer and consecutively high-energy ball milled (HEBM) in a Spex 8000D mill (SpexCertPrep, NJ, USA). The HEBM was performed using tungsten carbide lined stainless steel jars and 11.2 mm tungsten carbide milling media (with a ball-to-powder ratio of ~ 4.5:1) for 50 min with 1 wt% (0.05 g) stearic acid as lubricant. The as-milled powder mixtures were subsequently loaded into 10 mm graphite dies lined with graphite foils in batches of 2.5 g. The HEBM and subsequent loading/handling of as-milled powder mixtures were conducted in an argon atmosphere with $O_2 < 10$ ppm to prevent oxidation.

Subsequently, the specimens were consolidated into dense pellets via SPS in vacuum ($10^{-2}$ Torr) using a Thermal Technologies 3000 series SPS (CA, USA). During the SPS sintering process, specimens were first heated to 1400 °C at a rate of 100 °C/min with a constant pressure of 10 MPa. The temperature was then further raised to 1700 °C at 30°C/min, and subsequently maintained isothermally for 10 min for final densification. Note that binary tetraborides $LaB_4$ and $YbB_4$ would dissociate at ~1800 °C [36, 41]. At the same time, the pressure was raised to 50 MPa at 5 MPa/min. Finally, the sintered specimen was left inside the SPS machine (powered off) to cool down to room temperature with the pressure released back to 10 MPa.

The sintered specimens were first ground to remove carbon contamination on the surface from SPS graphite tooling, and subsequently polished for further characterizations. X-ray diffraction (XRD) characterizations were performed at 30 kV and 15 mA with Cu Kα radiation on a Rigaku Miniflex diffractometer. Scanning electron microscopy (SEM), energy dispersive X-ray spectroscopy (EDS), and electron backscatter diffraction (EBSD) were conducted on a Thermo-Fisher (FEI) Apreo microscope equipped with an Oxford N-Max$^N$ EDS detector and an Oxford Symmetry EBSD detector.

Aberration-corrected scanning transmission electron microscopy (AC-STEM) and nanoscale EDS were conducted on a JEOL 300CF microscope operated at 300 kV to characterize the crystal nanostructure and nanoscale elemental distributions. The TEM foil was prepared using Thermo-Fisher (FEI) Scios focused ion beam/scanning electron microscope (FIB/SEM).

Specimens densities were measured via the Archimedes method. The theoretical densities were calculated based on the nominal compositions and the lattice parameters determined from the unit cell refinement of the XRD patterns.

Vickers microhardness tests were performed on a LECO hardness testing machine with a diamond microindentor at a standard loading force of 9.8 N (1 kgf) and hold time of 15 s, abiding to the ASTM Standard C1327. For each specimen, over 30 indentation tests were carried out at different locations to minimize the microstructural and grain boundary effects and ensure the statistical validity.

## 3. Results and Discussion

After sintering, all six specimens (listed in Table 1) demonstrate largely single $UB_4$-prototyped tetragonal HETB phases without detectable oxide impurities based on the XRD patterns shown in Fig. 2. The calculated XRD patterns based on the nominal compositions and measured lattice parameters from XRD (Table 1) for all sintered specimens are displayed in Supplementary Fig. S1 for comparison. It can be clearly observed that HETB structures represent the primary phases in all specimens after the in-situ reactive SPS. Nevertheless, XRD patterns reveal small amounts of $CaB_6$-typed hexaboride secondary phases (Space group: Pm-3m, No. 221) in HETB1, HETB2, HETB3, and HETB6. The presence of this boron-rich secondary phase is likely due to the existence of native oxides that alter the metal-to-boron ratios. Prior studies of fabricating high-entropy diborides [31, 33] and monoborides [10] suggest the requirement of extra boron additions to facilitate the oxide removal. Thus, stoichiometric amounts of boron and metal precursors could result in deficient metals (or excessive boron) in this case of HETB specimens, thereby promoting the formation of minor boron-rich hexaboride impurities. Moreover, some binary tetraborides involved in this study (viz. $LaB_4$, $PrB_4$, $NdB_4$, $SmB_4$, and $YbB_4$) are known to dissociate into hexaboride and metal-rich phases (with the nominal reaction: $3MB_4 \rightarrow 2MB_6 + M$, where M = La, Pr, Nd, Sm, and Yb) at elevated temperatures above 1800 °C before melting [36, 41]. Thus, the existence of hexaboride phases may relate to this decomposition as well. Careful examination of the XRD pattern also suggests the possible existence of trace amount of hexaboride in HETB4 (with a peak barely observable at 2θ of ~31°). However, no XRD detectable secondary phase is observed in HETB5: $(Y_{0.2}Nd_{0.2}Sm_{0.2}Gd_{0.2}Tb_{0.2})B_4$.

Measured densities obtained by the Archimedes method indicate that all specimens are highly dense with >98% relative densities (ignoring the influence of minor hexaborides as their densities are slightly lower than the tetraborides). The high relative densities can be directly confirmed by SEM micrographs on polished surfaces (Supplementary Fig. S2), which show that the black spots involving porosities and/or extra boron are merely 1-2 vol%.

Both the measured lattice parameters from XRD unit cell refinements and the averaged lattice parameters calculated by the rule of mixture (RoM) of individual binary tetraborides are tabulated in Table 1. Overall, the measured lattice parameters agree with RoM values with <0.5% differences for all six cases.

Interestingly, a careful examination of Table 1 reveals that the lattice parameters $c$'s measured by XRD are generally larger than those predicted by RoM (with the only exception of HETB1), but the measured lattice parameters $a$'s are always smaller than the RoM predictions. This interesting observation can be justified intuitively based on the quasi-layered crystal structure shown in Fig. 1 and discussed above. Here, the parameter $c$ depends more on large cations (vs. the average of large and small cations) with more rigid 2D boron layers on (001) plane, but less boron-boron covalent bonds along the [001] direction. Fig. 1(b) shows a schematic illustration of different densities of boron-boron covalent bonds within the (001) (i.e. the $a$-$a$) plane and along [001] (i.e. the $c$ axis) in HETBs. Furthermore, an early study [42] also suggests that the tetraboride is much less rigid along [001] than the cubic hexaboride along [100]. This more rigid (001) plane and less rigid [001] direction in HETBs leads to a scenario akin to high-entropy diborides [11, 43] with a similar observation of an expanded lattice parameter $c$ from the RoM averages.

The homogenous elemental distributions of metal cations for all specimens are verified by SEM-EDS elemental mapping (Fig. 3(c) and Supplementary Fig. S4). STEM-EDS elemental mapping was further conducted for the single-phase HETB5: $(Y_{0.2}Nd_{0.2}Sm_{0.2}Gd_{0.2}Tb_{0.2})B_4$ at nanoscale, as shown in Fig. 3(b). Moreover, AC-STEM high-angle annular dark-field (HAADF) imaging at a high magnification shows the uniform high-entropy (solid-solution) $UB_4$-prototyped tetraboride structure at atomic scale (Fig. 3(a)). Additional STEM micrographs on the same specimen at different magnifications and the corresponding fast Fourier transform (FFT) diffraction pattern of the crystal grain can be found in Supplementary Fig. S3. The combination of the characterization techniques confirms the formation of homogeneous high-entropy solid solutions.

SEM-EDS quantitative analyses are used to measure the cation compositions of all six specimens, and the results are listed in Table 1. The compositions measured by EDS differ from the nominal equimolar compositions by only 1-3 at%, which are within the typical EDS measurement errors. Thus, the nominal compositions are confirmed and adopted for the discussion in this study.

Combining the above results from XRD, SEM, AC-STEM, and EDS, it can be concluded that a new class of HETBs of the $UB_4$-prototyped tetragonal structure has been successfully synthesized. Thus, this study has further expanded the family of HECs to include a new and unique (highly anisotropic, tetragonal, and quasi-layered) crystal structure, as shown in Fig. 1. These HETBs are the second class of rare earth HEBs (after rare earth high-entropy hexaborides [26, 27]). Since the prior reported high-entropy hexaborides are prepared by thermal reduction from rare earth metal oxides as the powders [27] or porous [26] materials, these HETBs epitomizes the first class of dense rare earth HEBs made in bulk form. Moreover, these HETBs the only equimolar (highly anisotropic) tetragonal HECs made, albeit the reported non-equimolar tetragonal high-entropy sulfides [44] as well as tetragonal distortions in some cubic high-entropy oxides [24, 45]). Perhaps more significantly, these tetragonal HETBs possess a unique, highly anisotropic (quasi-layered) crystal structure that is markedly distinct from all prior reported HEAs and HECs.

The grain sizes, crystal orientations, and textures of all sintered specimens have been determined by EBSD analyses conducted on polished specimen surfaces normal to the direction of the applied pressure and current during SPS. Normal direction inverse pole figure orientation maps for all HETB specimens are illustrated in Fig. 4, where grain size distributions are shown in the insets. The measured grain sizes are 3.70 ± 3.20 μm for HETB1, 4.75 ± 5.76 μm for HETB2, 4.00 ± 2.88 μm for HETB3, 4.22 ± 2.61 μm for HETB4, 3.98 ± 2.23 μm for HETB5, and 3.48 ± 1.52 μm for HETB6. Under the identical SPS sintering condition and final densification temperature of 1700 °C, all sintered specimens exhibit similar averaged grain sizes of ~3.5-5.0 μm. However, specimens HETB1 to HETB4 contain a small number of large grains as well as clusters of small grains, which is widely observed in borides prepared by in-situ metal-boron reactive SPS [33, 46-48]. Unlike hexagonal diborides synthesized via reactive SPS [33, 49], no noticeable texture has been detected in these sintered HETB specimens, which agrees with the consistency in relative peak intensities between the measured and calculated XRD patterns (Fig. 2 vs. Fig. S1).

Vickers microhardness tests have been carried out for all specimens at a standard indentation load of 9.8 N, and the results are listed in Table 1. The hardness of these tetragonal rare earth based HETBs are in the range of 13-15 GPa. There are only limited prior reports on the measured hardness of binary rare earth tetraborides; the hardness values of $YB_4$ and $GdB_4$ have been reported to be 27.94 and 18.63 GPa, respectively, without specifying the indentation load (presumably at a low load, given the high values) [50]. On the other hand, the hardness of most binary rare earth hexaborides are reported to be within the range of 22-28 GPa at much reduced indentation loads of 0.98 N or lower according to a recent review [37]. Note that tetraborides are less rigid along [001] than the cubic hexaborides with octahedra boron chains [42]; thus, the hardness of tetraboride should be generally lower than that of the hexaboride counterpart. The hardness of these rare earth based HETBs are lower than refractory metal high-entropy monoborides and diborides synthesized via the similar route [10, 33] and the superhard hexagonal $WB_4$ [51], which is not surprising given the different crystal structures and bonding characters.

4. **Conclusions**

Via in-situ metal-boron reactive SPS, high-entropy rare earth tetraborides of tetragonal $UB_4$-structure have been successfully synthesized for the first time. Specimens were sintered to >98% dense bulk pellets with homogenous elemental distributions confirmed by both microscale SEM-EDS and nanoscale STEM-EDS. The averaged grain sizes were determined to between 3.5-5 μm. The measured Vickers microhardness at 9.8 N are in the range of 13-15 GPa. One scientifically interesting observation is the general expansion (vs. compression) of the lattice parameter $c$ (vs. $a$) measured by XRD in comparison with the rule-of-mixture prediction, which can be explained by the anisotropic bonding character in the unique quasi-layered (highly anisotropic) HETB crystal structure.

In a broader context, this study expands the family of high-entropy ceramics via fabrication of a new class of tetragonal high-entropy borides, e.g., single-phase $(Y_{0.2}Nd_{0.2}Sm_{0.2}Gd_{0.2}Tb_{0.2})B_4$, in a unique quasi-layered crystal structure that is distinctly different from any other HEAs and HECs reported to date.

Acknowledgement: This work is supported by an Office of Naval Research MURI program (Grant No. N00014-15-1-2863). Q.Y. and J.L. also acknowledge partial support from the Air Force Office of Scientific Research (Grant No. FA9550-19-1-0327) for the STEM work.

**Supplementary Information** includes:
- Supplementary Figures S1-S4.
- Supplementary Table S1.
- Supplementary References.

**Table 1. Summary of the six specimens synthesized via in-situ metal-boron reactive SPS.** Measured cation compositions are obtained from SEM-EDS analyses. Lattice parameters are measured from XRD, while averaged lattice parameters represent the means of the lattice parameters of binary tetraborides (shown in Supplementary Table S1) calculated based on the rule of mixture. The measured densities are obtained via the Archimedes method, while the theoretical densities are calculated from the XRD measured lattice parameters and the nominal compositions. Grain sizes are measured form EBSD maps. Vickers microhardness values are measured at a standard indentation load of 9.8 N (1 kgf). One scientifically interesting observation is represented by the general expansion (vs. compression) of the lattice parameter $c$ (vs. $a$) measured by XRD in comparison with the rule-of-mixture prediction.

| Specimen | Nominal Composition | Measured Cation Composition | Measured Lattice Parameters $a$, $c$ by XRD (Å) | Averaged Lattice Parameters $a$, $c$ by RoM (Å) | Theoretical Density (g/cm$^3$) | Measured Density (g/cm$^3$) | Relative Density | Grain Size (μm) | Vickers Microhardness at 9.8 N (GPa) |
|---|---|---|---|---|---|---|---|---|---|
| HETB1 | $(Y_{0.2}Nd_{0.2}Sm_{0.2}Gd_{0.2}Er_{0.2})B_4$ | $(Y_{0.17}Nd_{0.18}Sm_{0.21}Gd_{0.22}Er_{0.22})B_4$ | 7.1066, 4.0416 | 7.1416, 4.0468 | 6.01 | 5.91 | 98.3% | 3.70 ± 3.20 | 14.1 ± 0.8 |
| HETB2 | $(Y_{0.2}Sm_{0.2}Gd_{0.2}Er_{0.2}Yb_{0.2})B_4$ | $(Y_{0.21}Sm_{0.19}Gd_{0.21}Er_{0.18}Yb_{0.21})B_4$ | 7.0911, 4.0356 | 7.1086, 4.0272 | 6.23 | 6.12 | 98.2% | 4.75 ± 5.76 | 13.3 ± 0.8 |
| HETB3 | $(Y_{0.2}La_{0.2}Pr_{0.2}Dy_{0.2}Er_{0.2})B_4$ | $(Y_{0.22}La_{0.18}Pr_{0.20}Dy_{0.23}Er_{0.17})B_4$ | 7.1255, 4.0619 | 7.1810, 4.0572 | 5.89 | 5.78 | 98.1% | 4.00 ± 2.88 | 14.5 ± 0.9 |
| HETB4 | $(Y_{0.2}Nd_{0.2}Gd_{0.2}Dy_{0.2}Er_{0.2})B_4$ | $(Y_{0.22}Nd_{0.18}Gd_{0.20}Dy_{0.21}Er_{0.19})B_4$ | 7.1142, 4.0479 | 7.1110, 4.0284 | 6.08 | 6.02 | 99.0% | 4.22 ± 2.61 | 14.4 ± 0.7 |
| HETB5 | $(Y_{0.2}Nd_{0.2}Sm_{0.2}Gd_{0.2}Tb_{0.2})B_4$ | $(Y_{0.22}Nd_{0.20}Sm_{0.20}Gd_{0.19}Tb_{0.19})B_4$ | 7.1331, 4.0641 | 7.1514, 4.0552 | 5.90 | 5.85 | 99.2% | 3.98 ± 2.23 | 13.9 ± 0.8 |
| HETB6 | $(Nd_{0.2}Sm_{0.2}Gd_{0.2}Tb_{0.2}Yb_{0.2})B_4$ | $(Nd_{0.18}Sm_{0.19}Gd_{0.23}Tb_{0.21}Yb_{0.19})B_4$ | 7.1254, 4.0628 | 7.1402, 4.0526 | 6.44 | 6.36 | 99.2% | 3.48 ± 1.52 | 14.1 ± 0.8 |

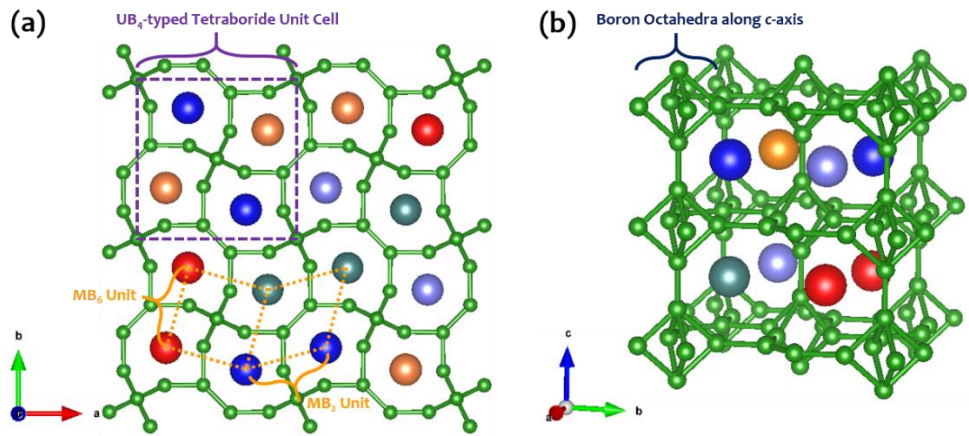

**Fig. 1.** Schematic illustration of atomic structure of the tetragonal high-entropy rare earth tetraboride (Space group: P4/mbm, No. 127) through **(a)** top view along *c*-axis and **(b)** perspective side view. The UB$_4$-prototyped tetraboride unit cell is indicated by the purple dash lines. Its structure can be interpreted as alternating (distorted) MB$_2$ and MB$_6$ units (each manifested by orange dotted lines respectively). Along *c*-axis, octahedra composed of six boron atoms are connected linearly. Notably, this tetragonal HETB crystal structure is "quasi-layered" with alternating covalently-bonded boron layers and "2D high-entropy" metal layers, thereby being highly anisotropic. However, unlike high-entropy diborides of the hexagonal AlB$_2$ structure [11], the boron layers are still linked by linear boron-boron chains along the *c* axis at a low density, and high-entropy metals are "caged" by covalently-bonded boron network or "isolated" from one another. Thus, the unique quasi-layered crystal structure of these TETBs is distinctly different from any other HEAs and HECs reported to date. Here, small green balls denote (covalently bonded) boron atoms, and colored large balls represent atoms of five different rare earth elements selected from Y, La, Pr, Nd, Sm, Gd, Tb, Dy, Er, and Yb.

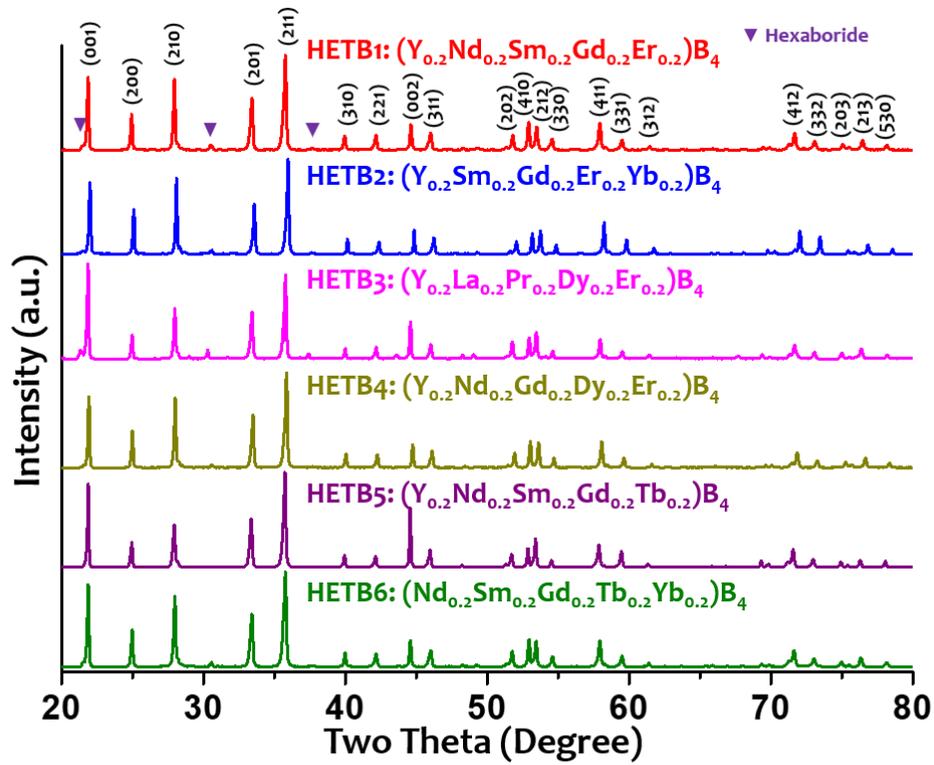

**Fig. 2.** XRD patterns of specimens HETB1 to HETB6 fabricated via in-situ metal-boron reactive SPS. All sintered specimens demonstrate largely a single $UB_4$-prototyped tetragonal phase without detectable oxide impurities. Small amounts of $CaB_6$-prototyped hexaboride secondary phases (Space group: Pm-3m, No. 221) are observed for some specimens, while no secondary phase is detected in HETB5 $(Y_{0.2}Nd_{0.2}Sm_{0.2}Gd_{0.2}Tb_{0.2})B_4$. Note that some minor diffraction peaks for $UB_4$-prototyped tetragonal phase are not indexed due to their low peak intensities and peak overlapping.

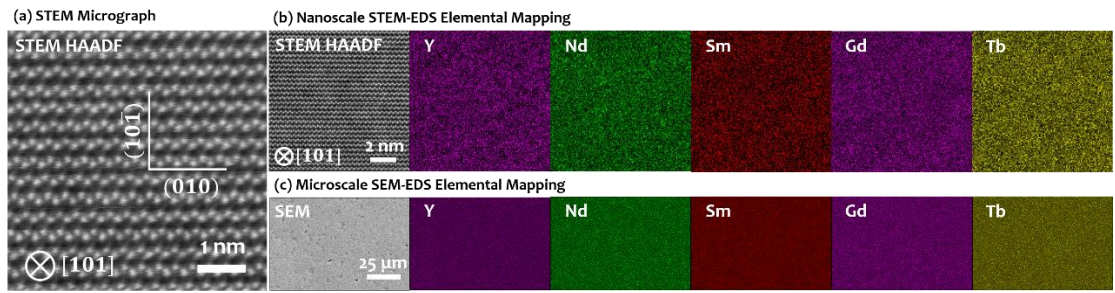

**Fig. 3.** (**a**) STEM HAADF image of the single-phase HETB5 $(Y_{0.2}Nd_{0.2}Sm_{0.2}Gd_{0.2}Tb_{0.2})B_4$, viewed long the [101] zone axis. See Supplementary Fig. S3 for additional STEM micrographs at different magnifications and the FFT diffraction pattern. (**b**) STEM micrograph and corresponding EDS elemental maps at nanoscale. (**c**) SEM micrograph and corresponding EDS elemental maps at microscale. The compositions are homogenous at both nanoscale and microscale. See Supplementary Fig. S4 for additional SEM-EDS elemental maps of other specimens.

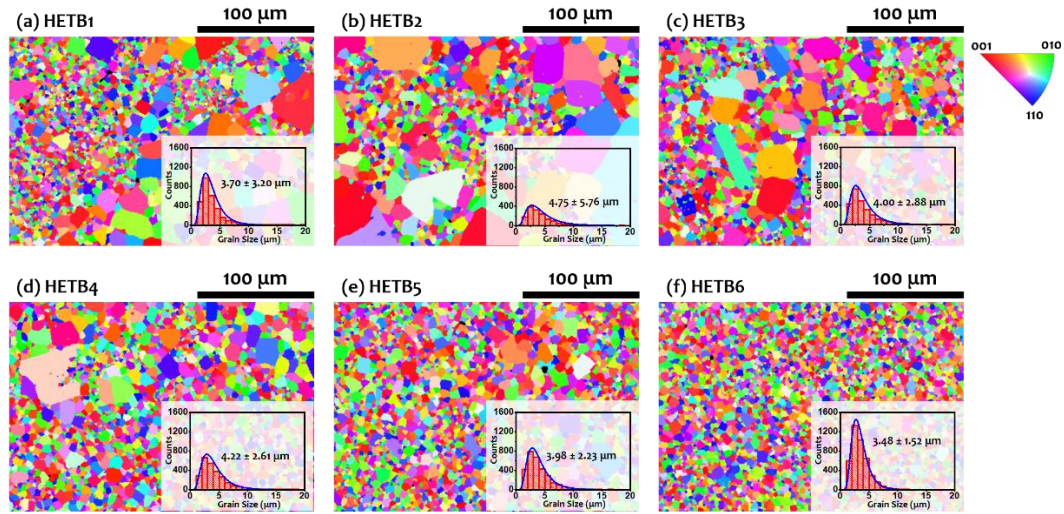

**Fig. 4.** EBSD normal direction inverse pole figure orientation maps for all six specimens **(a)** HETB1, **(b)** HETB2, **(c)** HETB3, **(d)** HETB4, **(e)** HETB5, and **(f)** HETB6. Specimens HETB1 to HETB4 show abnormal grain growth to different extents. All specimens contain some clusters of small grains, which are commonly observed for in-situ metal-boron reactive sintered specimens. Insets show the grain-size distributions.